# Integrated photonic multigrid solver for partial differential equations

*Timoteo Lee,*[1,2] *Frank Brückerhoff-Plückelmann,*[1] *Jelle Dijkstra,*[1] *Jan M. Pawlowski,*[2] *and Wolfram Pernice*[1]*

[1] University of Heidelberg, Kirchhoff-Institut für Physik, Im Neuenheimer Feld 227, 69120 Heidelberg, Germany.

[2] University of Heidelberg, Institut für Theoretische Physik, Philosophenweg 16, 69120 Heidelberg, Germany.

*Correspondence to: wolfram.pernice@kip.uni-heidelberg.de

**Solving partial differential equations is crucial to analysing and predicting complex, large-scale physical systems but pushes conventional high-performance computers to their limits. Application specific photonic processors are an exciting computing paradigm for building efficient, ultrafast hardware accelerators. Here, we investigate the synergy between multigrid based partial differential equations solvers and low latency photonic matrix vector multipliers. We propose a mixed-precision photonic multigrid solver, that offloads the computationally demanding smoothening procedure to the optical domain. We test our approach on an integrated photonic accelerator operating at 2 GSPS solving a Poisson and Schrödinger equation. By offloading the smoothening operation to the photonic system, we can reduce the digital operation by more than 80%. Finally, we show that the photonic multigrid solver potentially reduces digital operations by up to 97 % in lattice quantum chromodynamics (LQCD) calculations, enabling an order-of-magnitude gain in computational speed and efficiency.**

**Introduction**

Solving partial differential equations (PDEs) fast and efficiently is fundamental to all natural sciences and engineering, ranging from fluid mechanics to atmospheric modelling and groundwater simulation, to fundamental physics problems such as those found in lattice quantum chromodynamics[1–4]. As illustrated in **Fig. 1a** and **Fig. 1b,** we obtain numerical solutions by discretizing the system leading to systems of linear equations

$$Ax = b, \qquad (1)$$

which account for up to 99% of the computational cost of large-scale simulations[2]. High-precision solutions of the PDEs require a finer discretization of the system, but this leads to very large systems of linear equations with billions of unknowns that require massive amounts of computational power.

In the last decades, the steady increase of digital computational power has facilitated solving larger and larger linear systems. However, as Moore's Law starts to slow down and with the end of Dennard's scaling, new computing paradigms and application specific hardware are crucial to sustain the needed performance growth[5,6]. Specialized ultra-low latency photonic processors provide an interesting alternative to modern general-purpose parallel digital hardware for the sequential PDE solvers[7–9]. Photons, unlike electrons, can transmit information without being subject to ohmic losses and capacitive charging, enabling high-speed processing[10,11]. This technology excels at matrix operations deploying broadcast and weight architectures which enable matrix multiplications in a single clock cycle, offering speedups of more than two orders of magnitude in comparison to conventional graphic processing units[7].

Mixed-precision methods, where most of the calculation is done in low-precision, and high-precision correction steps are used sparsely to ensure convergence with high accuracy, are essential to exploit the strengths of photonic accelerators due to their inaccurate analog nature[12,13]. For example, enhancing photonic accelerated linear solvers with the residual iteration method allows to accurately solve a broad range of problems[12–14]. In this context, photonic computing complements rather than replaces its digital counterpart.

In this paper, we investigate mixed-precision photonic solvers based on multigrid methods (MPPMG) as a promising path to accelerate large-scale calculations. During computation, we offload matrix multiplications with constant weights to the photonic accelerators, enabling ultra-low latency in-memory computation. We test the hybrid system by solving a Schrödinger and Poisson equation. Finally, we explore the usage of the MPPCG solver for lattice quantum chromodynamics calculations.

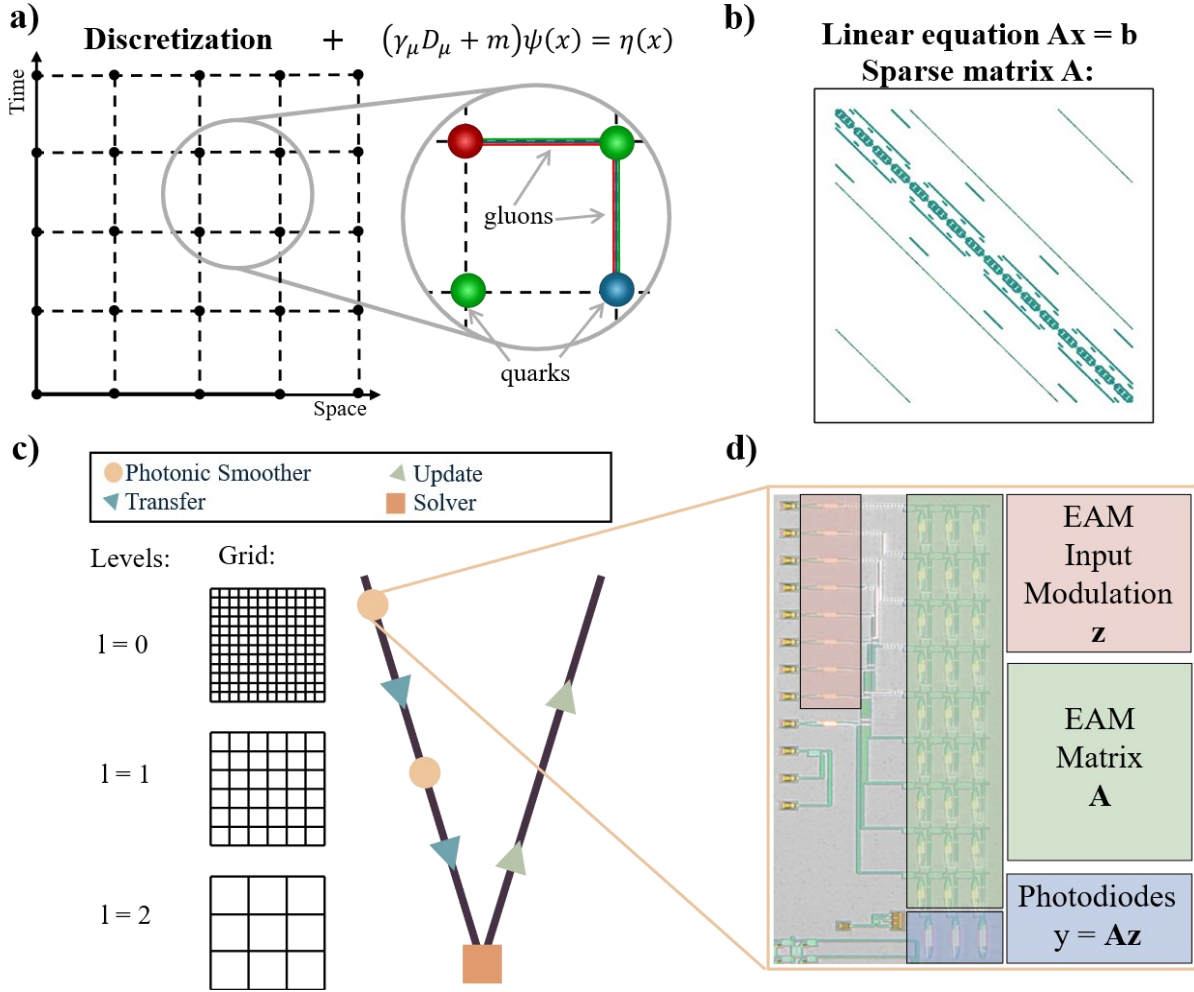

**Fig. 1. Photonic Partial Differential Equation Solver. a)** PDEs are essential building blocks for modeling our world. To compute them numerically, the theoretical models are solved in a discretized version of the system, for example, the physics of nucleons is described on a discretized space-time in lattice quantum chromodynamic (LQCD) calculations. **b)** Solving sparse system of linear equations, like the LQCD matrix found on the SuiteSparse matrix collection[15], is the bottleneck of many large-scale calculations. **c)** Multigrid methods are a powerful tool to solve linear equations. They rely on representing the system on coarser grids to solve the problem at different scales. We use an asymmetrical V-cycle, where we start from the finest grid (level 0), transfer the problem to coarser grids (levels 1 and 2), and improve the solution on the finer grids via the update steps. **d)** We offload the smoothening operations to an integrated photonic processor designed for analog in-memory computing. Electro absorption modulators (EAMs) encode the input vectors in the amplitude of light pulses, and the matrix elements are stored in tunable absorbers. Photodetectors read out the result of the matrix multiplication.

**Mixed-precision photonic solvers based on multigrid methods**

Multigrid methods eliminate the error, defined in the Methods Section, across all frequencies by operating on coarser representations of the system as sketched in **Fig. 1c**. In this paper, we offload the computationally demanding smoothing operation to an integrated photonic processor, shown in **Fig. 1d**, consisting of a 9x3 crossbar array storing the matrix weights for in-memory computing and highspeed electro absorption modulators and photodetectors to convert the input vectors to the optical domain and readout the results. The full processor is interfaced by DACs/ADCs running at 2 GS/s, enabling highspeed matrix multiplications[9]. In this configuration, as long as the photonic smoother eliminates the relevant high-frequency components well enough, the multigrid solver damps the low-frequency error components as illustrated in **Fig. 2a**. We improve the robustness of the photonic multigrid solvers against analog noise by using them as preconditioners for Krylov solvers, like the conjugate gradient (CG) solver, as shown in **Fig. 2b**. In this mixed-precision configuration, we solve the system of linear equations

$$MAx = A'x = b' = Mb. \qquad (2)$$

where M is the photonic multigrid preconditioner. The preconditioner helps the main solver converge with considerably less iterations as illustrated in **Fig. 2c**, by dampening the low-frequency components that are hard to eliminate for the main solver.

In this paper, we construct the multigrid solvers with the help of the Python library PyAMG[16], and we choose the Richardson smoother described by

$$x_{k+1} = x_k + \omega(b - Ax_k), \qquad (3)$$

where $\omega$ is the inverse of the approximate spectral radius of $A$. We plug in the resulting multigrid preconditioner directly to the different solvers implemented in SciPy[17] and stabilize the solver by extending it with the residual iteration, which resets the accumulation of rounding errors. For estimating the performance gain both, in terms of speed and energy-efficiency, we use the metric

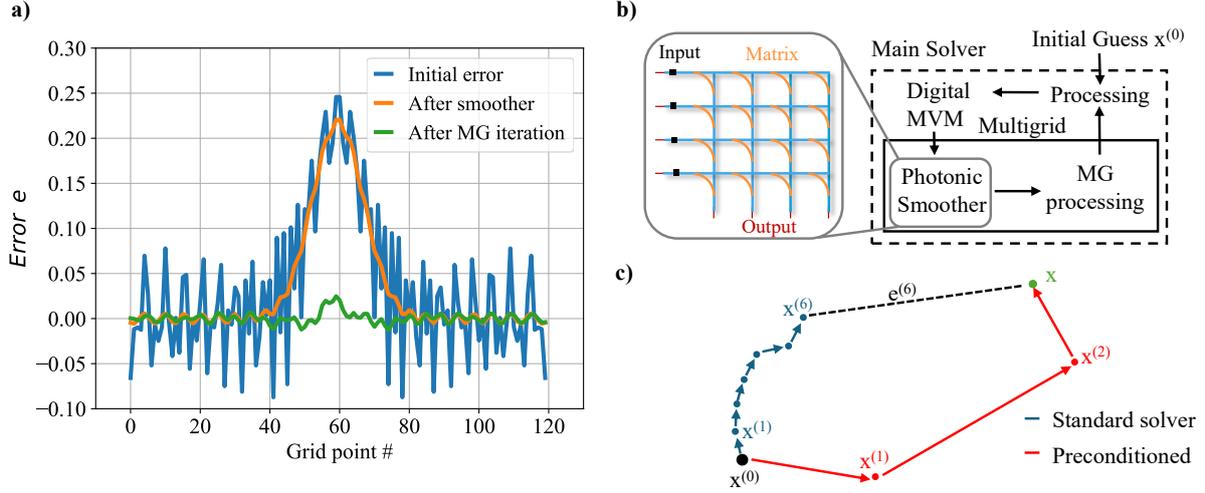

**Fig. 2. Mixed-precision Photonic Multigrid Method. a)** The initial error as defined in Eq. (5) has different frequency components, and the photonic smoother acts like a low-pass filter smoothening the error. The full multigrid iteration dampens the low-frequency components, which are hard to eliminate for the main solver. **b)** At each iteration, the main solver performs a digital matrix-vector multiplication (MVM), a multigrid iteration with the computationally demanding smoothening procedure offloaded to the optical domain, and some processing steps involving scalar and vector operations. **c)** The standard linear solver moves closer to the solution at each iteration. The fast photonic multigrid preconditioner helps the digital solver find better search directions towards the solution, accelerating the convergence.

$$G = \frac{N_d^{(S)}}{N_d^{(MP)}}, \qquad (4)$$

where $N_d^{(S)}$ and $N_d^{(MP)}$ are the number of double precision operations required by the double-precision solver and the mixed-precision solver, respectively. This value represents the upper limit of the performance gain of the proposed mixed-precision photonic solver. The details of the solvers and the metric are explained in the Methods Section. It is important to note that we make the approximation that the MVMs are the only relevant operations to the computation cost, and that the cost of the solver in the coarsest grid is negligible, which is the default approximation in literature[18,19]. Additionally, due to empirical observations, we choose to perform asymmetrical smoothening for the mixed-precision solvers like in the work by Bouwmeester et al.[20], where no post-smoothing is done. For a fair comparison, we perform four photonic pre-smoothening steps for the hybrid solvers, and two pre-smoothening and two post-smoothening for the digital solvers, keeping the total number of smoothening steps equal. Finally, due to the complexity of the lattice quantum chromodynamic calculation, we do not perform these calculations on our current experimental setup, and we introduce two emulated hybrid solvers to study the potential of these hybrid solvers on this problem. One emulates our experimental setup by performing noisy digital 8-bit operations, which is explained in the Methods section, and the other emulates an ideal 8-bit processor.

**Eigenvalue problem: The quantum quartic anharmonic oscillator**

Many scientific and engineering fields rely on efficient algorithms for solving eigenvalue problems, from quantum chemistry and materials science to imaging, data mining, and structural analysis. The locally optimal block preconditioned conjugate gradient (LOBPCG) method, which requires a good preconditioner to work efficiently, is widely used for these problems[21–33]. Here, we test a mixed-precision photonic eigensolver on the quantum quartic anharmonic oscillator (QQAO), which does not have analytical solutions. This eigensolver consists of a LOBPCG solver with a smoothed aggregation multigrid preconditioner with three levels. We observe that the photonic eigensolver uses 80% less double-precision operations than the digital eigensolver for the QQAO to converge to the same accuracy, see **Fig. 3c**. With a potential performance gain of 5-7 described by Eq. (4), the proposed solver has the potential to meaningfully integrate photonics to solve large-scale eigenvalue problems.

**Poisson problem: The parallel capacitor**

Algorithms for solving Poisson-like equations, also known as Poisson solvers, find applications in astrophysics, chemistry, mechanics, electromagnetics, statistics, and image processing[34,35]. The resulting systems of linear equations involve billions to trillions of unknowns, necessitating highly efficient numerical methods and scalable hardware architectures. Here, we calculate the electric field of a parallel-plate capacitor (PC) using a mixed-precision solver that consists of a conjugate gradient (CG) solver with a smoothed aggregation multigrid preconditioner with two levels combined with the residual iteration (RI). In this test, the proposed mixed-precision photonic solver reduces the double-precision operations by 60% and this could increase to 80% with higher-accuracy photonic MVMs, as can be seen in **Fig. 3d**. This leads to an estimated performance gain of 2.5-5, showing that these mixed-precision photonic Poisson solvers could serve as the starting point for building domain-specific hardware for many applications.

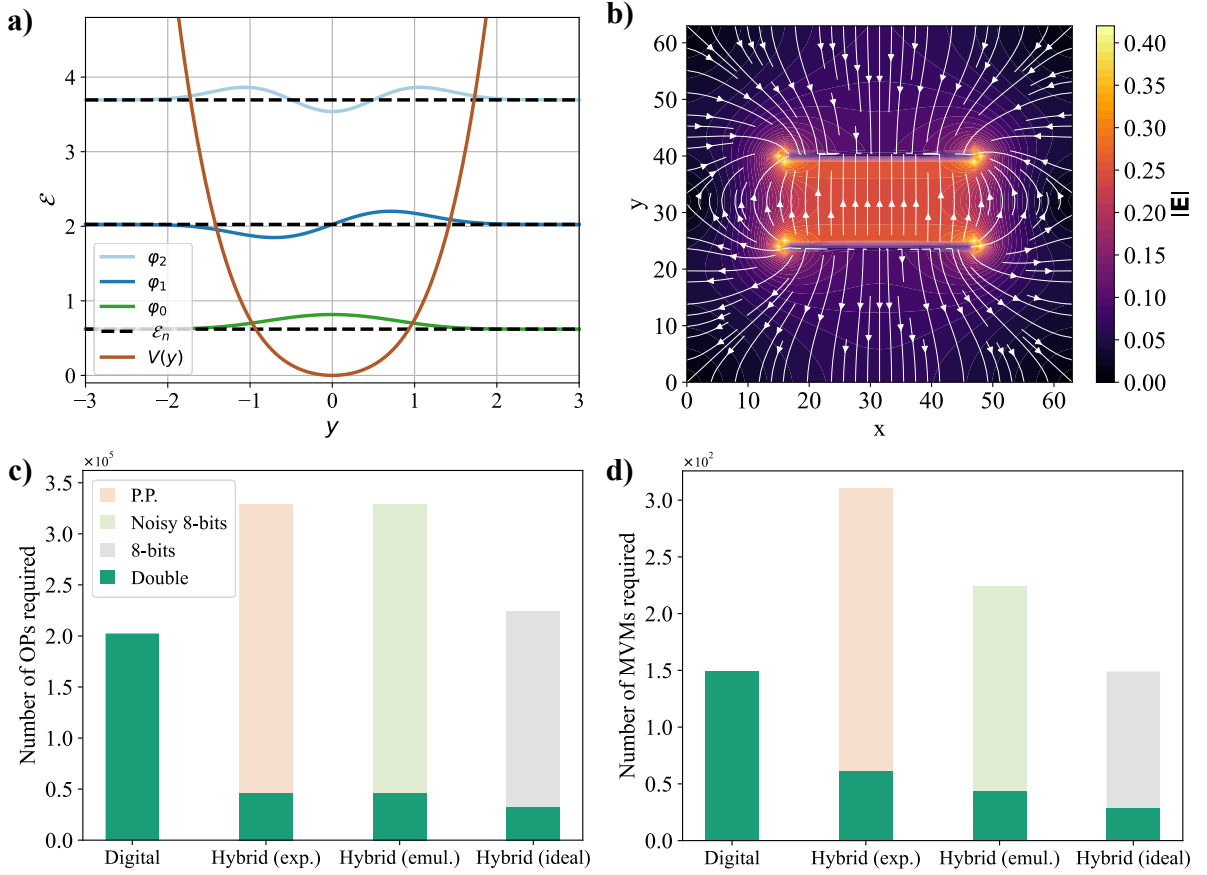

**Fig. 3. Photonically Accelerated PDE Solutions. a)** and **b)** We use the photonic in-memory processor (P.P.) to find the eigen-solutions of the quantum quartic anharmonic oscillator and the electric field of a parallel capacitor using the proposed mixed-precision photonic solver. **c)** and **d)** For the different solvers we compare the required total number of operations to converge with $\|r\| = 10^{-10}$ accuracy. The mixed-precision photonic solvers reduce the number of required high precision digital operation by 80% and 60%, respectively. In general, increasing the accuracy of the low precision smoother accelerates to overall convergence. The photonic processor behaves similar to the emulation of the system denoted by noisy 8-bits.

**Adaptive Multigrid Methods: Lattice Quantum Chromodynamics**

Quantum chromodynamics (QCD) describes the interactions between quarks and gluons as illustrated in **Fig. 4a**. Lattice field theory provides the standard numerical framework to study QCD non-perturbatively. However, lattice QCD (LQCD) calculations are computationally intensive, consuming a substantial share of global supercomputing resources[1]. In these calculations, solving systems of linear equations such as (1) accounts for most of the computation time, and the LQCD community has long pursued specialized hardware to accelerate these computations[12,36–38]. Adaptive multigrid methods revolutionized LQCD calculations due to their ability to deal with the phenomenon known as the critical slowing down[1,19]. The adaptive part can be seen as the "training phase" of the algorithm, where it learns to construct a multigrid solver able to deal with the critical slowing down. Then, we can use the resulting solver to find the solutions of the systems of linear equations with different right-hand sides (RHS) $b$, which can be seen as the "inference phase". The computational cost of both the training and inference phases is dominated by the smoother as illustrated in **Fig. 4b**. Future LQCD calculations need computers at least one order of magnitude quicker than the Exaflop computers highlighting the need to co-optimize the next generation of hardware and algorithms[39].

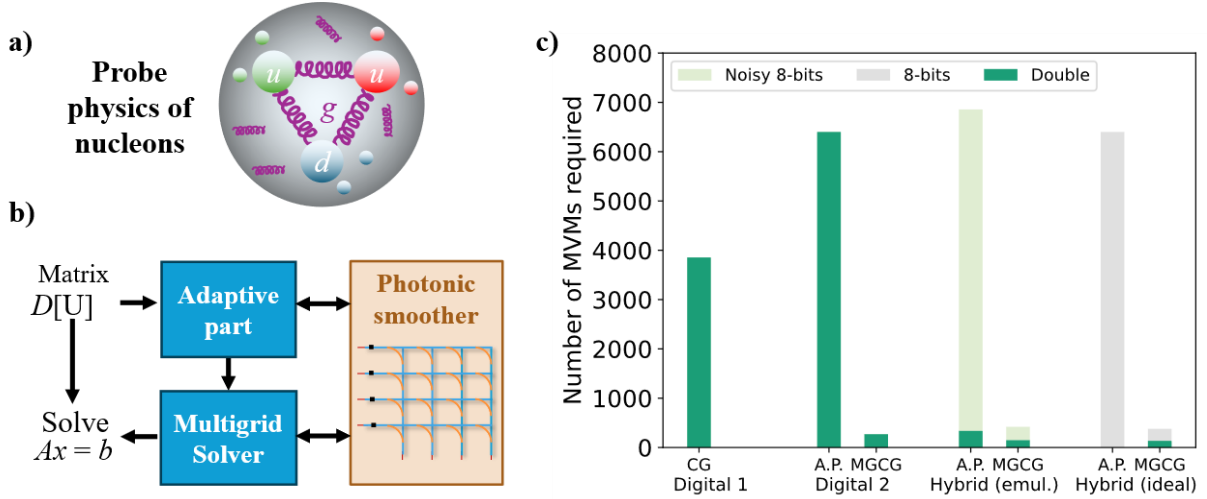

**Fig. 4. Lattice QCD calculations. a)** Quantum chromodynamics is a theory describing the physics of nucleons. **b)** After the discretization described in **Fig. 1a**, we get a matrix $D$, like in Eq. (13), describing the dynamics of quarks on a gluon background represented by the gauge links configuration U. To solve the resulting linear equation, we first "train" the algorithm during the adaptive part using the photonic smoother. **c)** The number of MVMs required for the different methods to converge for a single RHS is plotted. In digital 1, we use a standard conjugate gradient (CG) method with double precision, and in the rest, we use the multigrid preconditioned CG (MGCG) method with the adaptive part. We can see that the total number of operations required for convergence is similar for digital 2 and the hybrid solvers highlighting the robustness of the method to noise. Comparing digital 1 with the hybrid solvers, we observe a reduction of high-precision digital operations of 86-97%.

We investigate the potential of our mixed-precision photonic solver emulating calculations for the Wilson-Dirac operator of the 2D U(1) Schwinger model similar to the work done by Brannick et al.[19]. We solve 100 linear equations using the same mixed-precision solver as the Poisson problem with the addition of the adaptive part, which consists of the photonic smoother and the RI method. All the details are included in the Methods section. In **Fig. 4c**, we show the results for different solvers. To estimate the performance gains of the proposed solver for a single RHS, we compare the performance of the standard solver (digital 1) with the emulated hybrid solvers. The computational cost of the hybrid solvers is the combined cost of the adaptive part and solving a single RHS, and we observe a reduction of high-precision digital operations by 87-97%, which is equivalent to a maximum performance gain both in terms of speed and energy consumption of 8-30.

**Discussion**

Mixed-precision photonic PDE solvers based on multigrid methods (MPPCG) present an exciting approach to accelerate large-scale simulations with neuromorphic photonic processors. MPPCG solvers enable offloading the computationally demanding smoothening procedure to the optical domain leading to a potential one order of magnitude improvement in terms of speed and energy efficiency compared to fully digital solvers. Our results show that MPPCG solvers can have an especially large impact on the computationally hungry LQCD calculations. Current LQCD calculations based on multigrid methods struggle at increasing performance on modern general-purpose parallel hardware due to the sequential nature of the algorithms, and are limited by data movement due to the low-arithmetic intensity[1,40]. Specialized analog hardware provides additional tools with potential to solve these problems if meaningfully combined with digital hardware, for example, by leveraging low latency in-memory computing.

Due to the robustness with respect to noise, our mixed-precision approach is also compatible with other analog computing systems. While featuring less bandwidth than its photonic counterpart, electronic analog computing is appealing for dense problems. Due to the smaller component size, more than four million analog in-memory computing cells can be integrated on-chip with an overall compute computation accuracy of 3-4 bits[41]. Staying in the photonic domain, increasing the total number of matrix values ultimately requires encoding in multiple degrees of freedom[11]. Free space photonics compute systems offer an intriguing trajectory for scaling by leveraging an additional dimension for weight encoding[42]. Since the efficiency of theses system can increase with size, this ultimately enables ultra efficient matrix processing with less than 1 photon per multiply and accumulate operation[43].

In conclusion, analog in-memory processors have the potential to provide transformative performance gains for large-scale calculations through multigrid methods. Utilizing the high bandwidth of photonic processors, we offload computational demanding linear operations to the physical domain with ultra-low latency, enabling a novel class of hybrid, mixed-precision solvers.

## Methods

### Error analysis

The error of the linear solution is defined as

$$e := x - x^{(i)}, \tag{5}$$

where the $x^{(i)}$ is the current guess and $x$ is the real solution. The error can be decomposed into oscillatory components which are the eigenvectors of the matrix A. In short, the error can be written as

$$e = \sum_{n=0}^{N} c_n v_n, \tag{6}$$

where $v_n$ are the eigenvectors of A.

### Experimental setup

We use a setup that includes an amplified spontaneous emission light source (Agilent 83438A), an incoherent optical 9x3 crossbar array fabricated by imec on a silicon-on-insulator (SOI) platform integrating electro-absorption modulators for input modulation and weighting and photodiodes for readout. We control the full system by a SoC-FPGA (Xilinx ZCU 216) featuring 4 GSPS DACs and 2 GSPS ADCs for fast optical processing[9]. The weights for in-memory computing are programmed via a slower multi-channel DAC. We round the matrix values to 4-bit precision to reduce the amount of slow weight changes, and the input values to 8-bit precision. Due to the intrinsic optical noise of the ASE light source, we repeat the calculations 256 times and take the average as the result of the calculations. With this setup, we achieve errors comparable to digital 4-5 bit precision fixed-point calculations. Using a noise free coherent light source removes the need for averaging[9].

### Digital Noisy Fixed-Point Operations

We use digital noisy fixed-point operations to mimic the analog nature of our experimental setup for the LQCD calculation by rescaling the input and matrix values to the range [-1,1] with 8-bit discretization, adding Gaussian noise with a standard deviation of 0.045, performing the matrix-vector multiplications, and rescaling accordingly.

**Residual Iteration Method**

Intuitively speaking, the residual iteration (RI) method resets the solver free from the accumulated error, bringing further numerical stability to the solver. The steps of this method are

1. Choose initial guess x
2. Calculate residual r = b – Ax
3. Approximate the solution of Ae = r with some iterations of the solver
4. Update x ← x + e
5. Repeat 2 to 4 until convergence with the desired accuracy

**Approximation of the spectral radius**

The spectral radius of A is the largest absolute value among its eigenvalues. We approximate this by

1. Sampling a random vector x.
2. Updating x ← Ax.
3. Normalizing x.
4. Repeating 2 and 3 N times in total.
5. Calculating $\rho = x^T Ax$.

**Estimation of the performance gain**

We introduce a metric to evaluate the performance of the proposed solver. The goal of a good mixed-precision photonic solver is to offload most of the calculations to the optical domain. We estimate the maximum possible performance gain by assuming that the speed and energy efficiency of specialized hardware based on PICs are infinitely better than those of their digital counterparts. This would imply that current technical challenges, like data movement, are solved. In this case, only the double-precision operations contribute to the computation time and the energy consumption leading to an effective performance gain in both speed and energy efficiency described by (4).

**The quantum quartic anharmonic oscillator**

We solve the Schrödinger Equation for the quantum quartic anharmonic oscillator (QQAO). The eigenvalue problem for this Hamiltonian is

$$\frac{1}{2}\left(\frac{\partial^2}{\partial y^2} - y^2 + 2\mu y^4\right)\varphi(y) = E\varphi(y). \tag{7}$$

We use the finite-difference method (FDM) to discretize y in the domain [-3,3] with 120 points separated by a distance of h = 0.05, and we get the matrix

$$H_{ij} = \frac{1}{2}\left(\frac{2\delta_{i,j} - \delta_{i,j+1} - \delta_{i,j-1}}{h^2} - \left(y_j^2 + 2\mu y_j^4\right)\delta_{i,j}\right), \tag{8}$$

where we choose $\mu = 0.25$.

**The parallel capacitor**

We solve a simple two-dimensional electromagnetic problem, namely, a parallel capacitor problem. The Poisson equation is

$$\nabla^2 u(y, z) = f(y, z) = \frac{\rho}{\epsilon}, \tag{9}$$

and we discretize this again using FDM. Then, the linear equations have the shape

$$-4u_{i,j} + u_{i+1,j} + u_{i-1,j} + u_{i,j+1} + u_{i,j-1} = h^2 f_{i,j}, \tag{10}$$

where we chose h = 1 for simplicity. To solve the parallel capacitor problem, we need to use the correct boundary conditions like in the work by Zaman[35]. This leads to $f_{i,j} = 0$, except at the two plates where we have $f_{i,j} = 2$ and $f_{i,j} = -2$ for the respective plates. Then, the electrical field is given by

$$E = -\nabla u(y, z). \tag{11}$$

**Lattice Quantum Chromodynamics**

We follow similar calculations to Brannick et al.[19]. We provide here all the necessary data to reproduce the results in **Fig. 4c**. We use the quench approximation to generate U(1) gauge link configurations on a 128 x 128 lattice with the action

$$S[U] = \beta \sum_{n\in\Lambda} \sum_{\mu<\nu} Re\, tr[\mathbb{1} - U_{\mu\nu}(n)], \tag{12}$$

with periodic boundary conditions and with $\beta = 6$. We generate the gauge links using a Metropolis-Hasting algorithm. The Dirac matrix has the form

$$D_{x,y}[U] = -\frac{1}{2}\sum_{\mu=1}^{2}\left[(1-\gamma_\mu)U_x^\mu \delta_{x+\hat{\mu},y} \right.$$
$$\left. +(1+\gamma_\mu)\left(U_{x-\hat{\mu}}^\mu\right)^\dagger \delta_{x-\hat{\mu},y}\right] + (2+m)\delta_{x,y}, \qquad (13)$$

where $\gamma_\mu$ are the Euclidean gamma matrices, $U$ are the gauge link configurations, and $m$ is the bare fermionic mass. We solve the normal equation with $DD^\dagger$ as the matrix $A$, where both matrix-vector multiplications are processed individually with the digital noisy operations described above. In **Fig. 4**, we solve linear equations with $m = -0.063$ for 10 different gauge fields each with 10 different random RHS with an accuracy defined by $\|r\| = 10^{-10}$.

**Adaptive Multigrid Method**

The adaptive part is important to compute the low modes of the matrix, which are particularly challenging in lattice QCD calculations. We calculate these low modes by sampling a random Gaussian initial guess $x$ and solving $Ax = 0$ by applying the smoother in (4) many times to $x$, saving the result, and repeating this process 8 times in total. We block the lattice into 4 x 4 blocks for building the coarse matrix. For the implementation, we build the aggregate operator based on the 4 x 4 blocks and use the calculated low-modes and the PyAMG library to construct a smoothed aggregation multigrid solver.

With the noisy 8-bit smoother, we require to use the residual iteration method to reach the accuracy required. With 8-bits, the RI method can improve the resulting multigrid solver, but it also works well enough without any high-precision residual iteration steps. We choose not to use the RI method for the 8-bits calculation, which explains the lack of double-precision operations in the adaptive part (A.P.) of Hybrid (ideal).

## Acknowledgements

This work is funded by the Deutsche Forschungsgemeinschaft (DFG, German Research Foundation) by the Collaborative Research Centre SFB 1225 - 273811115 (ISOQUANT) and Germany's Excellence Strategy EXC 2181/1 - 390900948 (the Heidelberg STRUCTURES Excellence Cluster). We further acknowledge funding for this work from the European Union's Horizon 2020 Research and Innovation Program, Darpa NAPSAC and ERC AdG PICNIC.

## Author contributions

- Conceptualization: TL, FB, JP, WP
- Methodology: TL, FP, JD
- Investigation: TL, FP
- Visualization: TL, FB, WP
- Funding acquisition: JP, WP
- Project administration: JP, WP
- Supervision: FB, JP, WP
- Writing – original draft: TL
- Writing – review & editing: All authors

## Competing interests

The authors declare that they have no competing interests.

## Additional information

Correspondence and requests for materials should be addressed to W.P.

## Data availability

All data are available in the main text